\documentclass{article}
\usepackage{spconf,amsmath,graphicx}
\usepackage[subtle]{savetrees}
\usepackage{hyperref}
\usepackage{cite} 
\usepackage{booktabs}
\usepackage{xcolor}
\usepackage{subcaption}
\usepackage{multirow}
\usepackage{makecell}
\usepackage{array}

\title{ATTENDANCE-GUIDED SPEAKER ENROLLMENT ACROSS MEETINGS}
\name{Minjae Lee, Hee-Soo Heo, Youngki Kwon, Bong-Jin Lee}
\address{NAVER Cloud Corporation, South Korea}

\begin{document}
\ninept
\maketitle

\begin{abstract}
Speaker-aware applications rely on enrolled voice profiles, yet collecting enrollment speech from every participant is impractical.
Meeting archives provide an alternative source of enrollment speech, comprising recordings of recurring participants paired with expected attendance lists.
We investigate whether expected attendance lists can be cross-referenced with diarized anonymous speaker clusters to identify reliable enrollment speech for recurring participants.
The resulting task, multi-participant automatic enrollment, must accommodate two types of attendance noise: a listed participant may not speak, and a speaker may be unlisted.
To this end, we introduce a benchmark for the task: a synthetic corpus calibrated to an in-house meeting corpus, paired with a contamination-sensitive evaluation protocol.
The reference method we introduce scores cross-session cluster chains against expected-attendance patterns and abstains when the evidence is insufficient.
We quantify the reference method's benchmark performance as a trade-off between enrollment success and contamination.
Applying synthetic-development thresholds unchanged to real meetings exposes a synthetic-to-real gap that remains an open challenge.
We will release the benchmark generator and evaluation code.
\end{abstract}


\section{INTRODUCTION}
\label{sec:intro}

Personalized speech applications such as meeting transcription with speaker attribution, voice-based services, and smart assistants commonly rely on identifying speakers against enrolled profiles~\cite{Snyder2018,malegaonkar2011performance,peri2023voxwatch,kanda21b_interspeech,sholokhov22_odyssey,Kinnunen2010}.
Research on such downstream tasks typically treats the profiles as given and clean.
However, in real-world applications, these profiles must first be constructed.
Constructing voice profiles via explicit enrollment requires a dedicated recording from every user.
Consequently, the collection effort increases with the size of the participant pool.
Meeting archives, the recordings that accumulate as meetings recur, offer an alternative: they may already contain speech from recurring participants.
Two sources of evidence support naming this speech: each session's recording can be diarized into anonymous speaker clusters~\cite{anguera2012speaker,park2022review}, and each session is associated with an expected attendance list.
Such input pairs are commonly available at no extra cost: recorded recurring meetings provide the audio, and scheduling systems provide each meeting's invitee list, compiled in advance and seldom reconciled with actual attendance.
Importantly, this list is session metadata, not a record of who attended or spoke: a listed name may stay silent, and an unlisted person may speak.
Even a logged attendance record would not close this gap, since attending a meeting is not the same as speaking in it.
If a participant's attendance pattern across sessions is distinctive --- for instance, if they are the only one attending a particular subset of sessions --- the clusters that recur in exactly those sessions likely belong to them, and target-speaker segments can be collected there as enrollment speech.
The resulting enrollment speech can be used to construct voice profiles for recognition in later sessions, without requiring dedicated enrollment recordings---a process we call \emph{attendance-guided speaker enrollment}.

This attendance-guided enrollment risks profile \emph{contamination}: speech from another speaker enters a target participant's enrollment profile and corrupts downstream decisions.
Because enrollment quality depends on segment reliability rather than quantity, the system must abstain rather than enroll doubtful speech; risk-sensitive deployments must operate at low contamination.
As we show, this regime has the least favorable success--contamination trade-off.
Attendance-list uncertainty further increases this risk because the lists represent expectations rather than observations: a list may include a regular member who is absent or silent (\emph{inclusion noise}), and may omit a guest who speaks (\emph{omission noise}).

Prior work introduced metadata-assisted enrollment of the recording user but provided neither a quantitative evaluation nor a public benchmark~\cite{cohen2018fully}.
External metadata and on-screen text have been used to assign names to diarization clusters~\cite{huijbregts11_interspeech,6949118}.
Household speaker recognition has formalized passive enrollment with unknown guests~\cite{sholokhov22_odyssey}.
However, none of these studies formulates or benchmarks the enrollment of multiple recurring participants from noisy session-level attendance lists.
Existing public meeting corpora~\cite{janin2003icsi,vinnikov2024notsofar,carletta2005ami,yu2022m2met,watanabe2020chime,Fu2021} are not directly suited to this task either: they do not jointly provide recurring speakers with consistent cross-session identities, sufficient recurrence and attendance variation, and independently collected expected attendance lists.
Our in-house corpus provides all three, including the independently collected attendance lists that public corpora lack.
Like those corpora, it also carries per-segment speaker annotations, used only to calibrate and evaluate the benchmark, not as method input.
However, the corpus is insufficient in scale for reliable development and evaluation on its own, and expanding it would entail the same costly speaker- and segment-level annotation.
We therefore construct a synthetic benchmark that reproduces the corpus's characteristics and serves as the experimental basis of this work.

This paper makes three contributions.
\textbf{(1)}~We formalize multi-participant automatic enrollment from meeting archives with expected attendance lists and specify a contamination-sensitive evaluation protocol that selects operating points on development data and reports contamination--success curves.
\textbf{(2)}~We introduce a synthetic benchmark whose statistics---including speaker recurrence, session composition, and attendance-noise rates---are calibrated to our in-house corpus, with four conditions isolating inclusion and omission noise.
\textbf{(3)}~We present a reference method with a single tunable decision threshold and analyze its transfer from synthetic development data to real meetings without threshold re-tuning, quantifying the synthetic-to-real gap.
To the best of our knowledge, this is the first benchmark for extracting named enrollment speech for multiple recurring meeting participants from noisy session-level attendance lists.

\section{TASK DEFINITION}
\label{sec:task}

\begin{figure*}[t!]
\centering
\includegraphics[width=\textwidth]{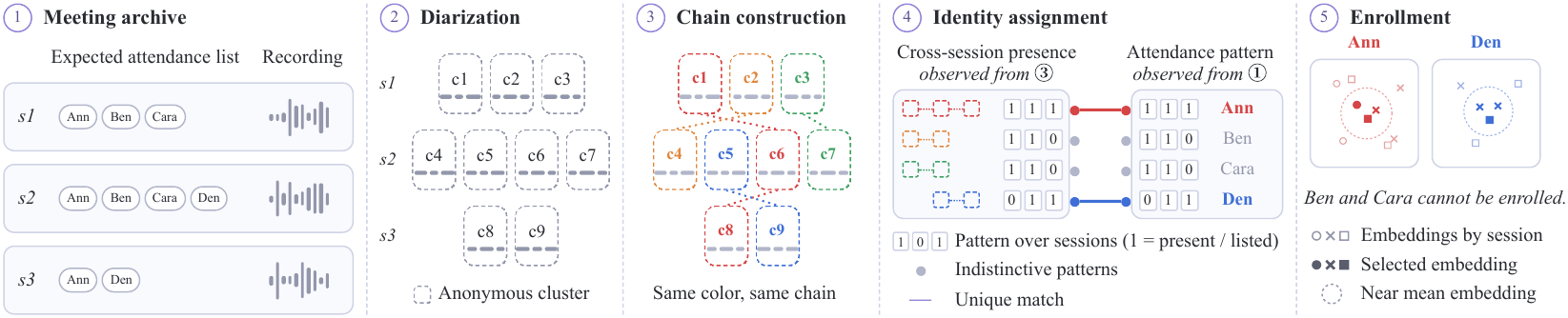}
\caption{Attendance-guided enrollment on a noise-free toy episode.
\textbf{(1)}~Each session provides an expected attendance list $L_s$ and a recording;
\textbf{(2)}~diarization yields anonymous single-speaker clusters;
\textbf{(3)}~cross-session voice similarity links clusters into chains;
\textbf{(4)}~a chain is assigned a name only when its cross-session acoustic evidence is sufficiently consistent with one participant's expected-attendance pattern---Ben and Cara share a pattern, so the method abstains on both; naming either would risk assigning the wrong speaker;
\textbf{(5)}~for each named participant, the segments closest to the assigned chain's pooled mean embedding are selected as enrollment speech.
For clarity, the toy is noise-free and omits the $m{=}3$ eligibility gate and the spurious per-session selections that the scoring stage (Section~\ref{sec:method}) suppresses; it illustrates the mechanism, not the exact decision rule.}
\label{fig:pipeline}
\vspace{-15pt}
\end{figure*}

This section formalizes attendance-guided speaker enrollment.
The task is to output uncontaminated target-speaker segments for each named participant, using only anonymous speaker clusters and the expected attendance lists.
This requires determining which clusters, if any, belong to each name and selecting reliable segments from the assigned clusters.
We first define the input archive and the cross-session evidence used for naming.
We then specify the diarization interface, attendance noise, speaker sets, evaluation metrics, and operating points.
An \emph{episode} is the archive-level unit processed jointly during enrollment.
It consists of recorded meeting sessions indexed by $s$, each providing an audio recording and an expected attendance list $L_s$ of participant names (Fig.~\ref{fig:pipeline}, stage~1).
The actual speaker names are not given.
A speaker-diarization front end partitions each recording into \emph{anonymous} single-speaker clusters (stage~2).
We exclude segments shorter than 4\,s as a fixed preprocessing step, since short segments yield less reliable embeddings~\cite{Kanagasundaram2011,Bhattacharya2017,jung2019short}.
Segment reliability matters more than quantity: even a few reliable segments suffice to enroll a speaker, and exhaustive labeling of every segment is secondary.
The enrollment method formalized and evaluated in this work takes only these clusters and the lists as input.
For each name, it either returns candidate segments believed to contain that person's speech or abstains.

\textbf{Cross-session evidence.}
The task cannot be resolved from a single session, whose list and clusters are unordered sets with no link between a name and a voice.
Naming instead rests on cross-session alignment: acoustic similarity links a speaker's recurring clusters into a chain (Fig.~\ref{fig:pipeline}, stage~3), and variation in the expected-attendance patterns is the only naming evidence (stage~4).
Participants with identical speech-presence patterns (e.g., Ben and Cara in Fig.~\ref{fig:pipeline}) stay indistinguishable from this evidence alone, motivating the evaluation population below.

\textbf{Diarization interface.}
To isolate attendance-driven identity association from diarization errors, we assume oracle \emph{within-session} speaker clusters and remove overlapping speech.
The enrollment method must infer which clusters across sessions belong to the same speaker.
These assumptions matter because every enrolled segment becomes trusted identity evidence.
Mixed speech and under-clustering violate cluster purity, whereas over-clustering divides one speaker's speech among multiple clusters.
Given only the expected attendance lists and anonymous clusters, such front-end errors are not generally distinguishable from attendance noise.
We therefore restrict the primary scope of this work to attendance noise and quantify the performance cost of relaxing the front-end assumptions diagnostically (Fig.~\ref{fig:curves-real}).

\textbf{Attendance noise and speaker sets.}\sloppy{}
Let $A_s$ be the set of named speakers observed speaking in session $s$, available only as ground truth for benchmark construction and evaluation.
Its mismatch with the expected attendance list $L_s$ produces two noise types that affect enrollment differently.
Inclusion --- a listed name that does not speak --- creates unsupported name--session associations and dilutes attendance evidence.
By contrast, omission --- a speaker who speaks but is not listed --- hides sessions in which the speaker was observed, weakening the cross-session pattern used for naming; a name's absence from a list thus provides contrastive, not conclusive, negative evidence.
We quantify these as two attendance-noise rates, $r_{\mathrm{inc}}=\sum_s |L_s\setminus A_s| / \sum_s |L_s|$ and $r_{\mathrm{om}}=\sum_s |A_s\setminus L_s| / \sum_s |A_s|$.
In the worst case, the inputs do not reveal whether a cluster is a listed participant's own speech, or whether that participant was silent and the cluster belongs to another (possibly unlisted) speaker, so any enrollment is ambiguous unless the method abstains.

To characterize the archive-level consequences of these two noise types, we distinguish four speaker sets.
The set $\mathcal{P}=\bigcup_s L_s$ of \emph{enrollment candidates} contains every name a method can output.
Inclusion noise can enlarge it, raise a candidate's observed list frequency, or introduce a name with no retained speech.
A speaker omitted from every list instead lies outside $\mathcal{P}$ --- an archive-level extreme of omission noise --- and is \emph{off-list}: unreachable by any list-based method and therefore excluded from the evaluation population.
The \emph{evaluation population}, the denominator of all evaluation metrics, contains speakers who appear on at least one attendance list, speak in at least one session, and whose \emph{speech-presence pattern} $a_p$ is unique within the episode.
This pattern $a_p$ is the binary vector over sessions in which a speaker is observed speaking, the ground-truth counterpart of the \emph{expected-attendance pattern} the method uses; it is used only to define the evaluation population and to score outputs, never as method input.
The remaining speakers are listed and speak but share their $a_p$ with another, and thus no attendance evidence can separate them; these \emph{non-evaluated} speakers are excluded from scoring, and their retained speech stays in the data as distractor clusters.

\textbf{Evaluation protocol.}
All evaluation metrics are computed over the evaluation population defined above, with their numerator and denominator counts pooled across episodes.
In this work, an enrollment method may output at most three segments per name, an evaluation-design choice matching the strong-success requirement.
The primary metrics are \emph{strong success} (enrolled with $\geq$3 target-speaker segments and no contamination) and \emph{contamination} (enrolled with any wrong-speaker segment).
The three-segment target permits averaging across multiple segments while remaining substantially less demanding than exhaustive extraction.
Strong success and contamination jointly measure whether sufficient uncontaminated enrollment material is extracted rather than whether the resulting profile improves a downstream recognizer.
We also define \emph{weak success} (enrolled with $\geq$1 target-speaker segment and no contamination).
Because a single target-speaker segment provides less enrollment evidence than the three-segment target, weak success serves only as a tie-breaker in threshold selection.
Enrollments under names outside the evaluation population are tracked separately and excluded from the primary metric numerators, although they may affect evaluated targets through name assignment.
Detailed counts accompany the released evaluation code.

For each contamination budget $\varepsilon$, the decision threshold is selected on the development split only, maximizing strong success subject to the budget.
Ties are resolved first by weak success and then by \emph{segment recall}, the fraction of target segments correctly enrolled.
The selected threshold is applied to the evaluation split without adjustment.
We report the resulting operating points over a budget grid, including the no-threshold limit (\emph{saturation}), together with the full contamination--success curve.
This operating-point and reporting procedure follows detection and selective-prediction methodology~\cite{martin1997det,el2010foundations,geifman2017selective}.
Because $\varepsilon$ bounds contamination only on the development split, the selected thresholds need not attain the same contamination on the evaluation set.

\section{BENCHMARK CONSTRUCTION}
\label{sec:benchmark}

\textbf{Real in-house corpus.}
We built a 63-session in-house meeting corpus of 84 speakers, 50 of whom form the evaluation population; the remaining speakers are off-list, retain no speech after filtering, or share a speech-presence pattern with another speaker.
The expected attendance lists $L_s$ are the meetings' invitation metadata, recorded independently of the audio; the corpus also carries per-segment speaker annotations produced separately from this metadata.
From these annotations we derive $A_s$, the speakers actually observed speaking, and form the oracle within-session clusters (overlapping speech removed; non-identifying labels excluded from targets but retained as anonymous distractor clusters).
The 4\,s minimum-duration criterion can leave a listed participant with no retained speech, indistinguishable at the method input from one who does not speak.
Table~\ref{tab:stats} reports the corpus's attendance statistics---speaker recurrence, session sizes, and inclusion/omission rates---that the synthetic benchmark below reproduces.

Speaker- and segment-annotated real meetings require substantial annotation effort.
Moreover, our corpus is too small for reliable development and evaluation splits.
We therefore select all decision thresholds and operating points on synthetic development data only, and retain the full corpus as a single-episode in-house evaluation/diagnostic set.

\textbf{Synthetic benchmark.}
We construct synthetic episodes from VoxBlink2~\cite{lin24j_interspeech}, calibrated to the in-house attendance statistics; no in-house recordings or speaker identities are included.
Each episode contains 63 sessions and 84 candidate identities, matched to the real corpus for direct comparability.
The session structure comprises recurring meeting series with partially overlapping participant groups, together with ad hoc sessions.
VoxBlink2 provides sufficient speakers for speaker-disjoint episodes.
Each speaker--session cluster draws segments from a single source recording, unused elsewhere, so cluster sizes follow the source's per-recording segment counts.
These clusters are oracle only with respect to the VoxBlink2 source labels.
Table~\ref{tab:stats} summarizes the resulting real--synthetic correspondence and the remaining cluster-size mismatch: real clusters have a markedly heavier upper tail than the generator reproduces.
Two of these statistics capture how distinguishable speakers are across sessions---the fraction of speaker pairs that never share a session, and the median number of sessions in which a pair's presence differs.
Larger values indicate more distinctive attendance patterns, and hence an easier association problem.

We apply attendance noise only to the metadata---removing listed names for omission and adding non-attending names for inclusion---while leaving the audio and speaker clusters unchanged.
We define four attendance-list conditions: the \emph{main} condition matches the real corpus's post-filter inclusion and omission rates; the \emph{clean} condition is an attendance-noise-free reference; and the \emph{inclusion-only} and \emph{omission-only} conditions isolate the two mechanisms.
All four conditions thus share the same embeddings, underlying speech, and cluster geometry; their evaluation populations can differ slightly because off-list targets are excluded per condition.
\begin{table}[t]
\centering
\caption{Real-corpus and synthetic-benchmark statistics.
Synthetic geometry and cluster statistics are shared across the attendance-list conditions, while the effective noise rates are those of the real-calibrated main condition.
Synthetic values are per-episode means, except for the percentage and median rows, which are pooled across episodes.}
\label{tab:stats}
\footnotesize
\setlength{\tabcolsep}{4pt}
\begin{tabular}{lcc}
\toprule
Statistic & Real & Synthetic \\
\midrule
Evaluation population & 50 & 58.3 \\
Speakers/session (median) & 6 & 6 \\
Never-co-occurring speaker pairs (\%) & 63.8 & 62.0 \\
Differing-presence sessions per pair (median) & 13 & 14 \\
Inclusion\,/\,omission rate $r_{\mathrm{inc}}\,/\,r_{\mathrm{om}}$ (\%) & 13.5\,/\,8.3 & 13.5\,/\,8.3 \\
Segments/cluster (median\,/\,p95) & 4.5\,/\,39 & 3\,/\,5 \\
\bottomrule
\end{tabular}
\end{table}
The development and evaluation splits contain 12 and 36 episodes, respectively, with disjoint speaker sets; the evaluation split has 2,099 targets in the main condition (2,072--2,115 across conditions), substantially more than the development split.

\section{REFERENCE METHOD}
\label{sec:method}

The reference method consists of four steps---cross-session chain construction (stage~3 of Fig.~\ref{fig:pipeline}), attendance-contrast scoring and global one-to-one assignment (stage~4), and enrollment-segment selection (stage~5)---where the figure illustrates the concept rather than the procedure specified below.
Only the decision threshold is tuned for each operating point.

\textbf{(1) Chain construction.}
A seed is any anonymous cluster in the episode, and one \emph{chain} is built per seed.
For each seed, the method visits every session and selects exactly one of its clusters: the candidate whose segments have the highest median pairwise cosine similarity to the seed's segments.
Because one cluster is selected in every session, some selections are spurious by construction: the method cannot know which sessions the seed's speaker truly attended.
The following scoring step suppresses such selections by contrasting listed against unlisted sessions.
Since all seeds are compared against the unchanged cluster pool, chain construction is independent of the seed-processing order.

\textbf{(2) Attendance-contrast scoring.}
In this scoring step, each participant is paired with each candidate chain.
For a participant--chain pair, the \emph{association score} is the chain's mean similarity over the participant's listed sessions minus its mean over the unlisted sessions.
A chain that corresponds to the participant is similar in the listed sessions and dissimilar in the unlisted ones, yielding a large positive score.
A chain similar across most sessions, or mainly in unlisted ones, scores near zero or negative.
Because both terms are means, the score does not grow with the number of listed sessions; an unnormalized sum would systematically favor frequently listed participants and preclude a common threshold across participants.

\textbf{(3) Global one-to-one assignment.}
Participants listed in at least $m{=}3$ sessions are matched to chains by the Hungarian algorithm~\cite{kuhn1955hungarian}.
This $m{=}3$ gate requires repeated list evidence before association; it is a fixed design constant, distinct from the three-segment enrollment budget.
An assigned participant--chain pair is rejected when its association score is below the decision threshold, selected on development data as in Section~\ref{sec:task}.
Each chain is assigned to at most one participant, although a cluster may appear in multiple chains.
The method also abstains on any eligible name whose expected-attendance pattern is shared by another eligible name, since the association score cannot separate them.
Their chains remain reserved in the assignment, but no enrollment is returned.
The $m{=}3$ gate leaves part of the evaluation population unreachable by construction.
We report this excluded fraction as the \emph{method-induced unreachable rate}, since other methods may attempt these participants.

\textbf{(4) Enrollment-segment selection.}
For each assigned participant (Fig.~\ref{fig:pipeline}, stage~5), the method retains only the chain's clusters from sessions in which that participant is listed, and pools their segment embeddings.
The retained embeddings are averaged into a mean embedding, and the three segments closest to it (by cosine similarity) form the enrollment profile, keeping only the most representative and reliable segments.
The selected segments need not come from different sessions.
This selection favors segments acoustically representative of the assigned chain.
It does not explicitly detect attendance noise and cannot recover evidence from sessions in which the participant was omitted from the list.


\section{EXPERIMENTS}
\label{sec:experiments}

\textbf{Setup.}
All experiments were conducted with a single pre-trained speaker embedding extractor, the large-margin fine-tuned ReDimNet2-B3~\cite{yakovlev24_interspeech,yakovlev2026redimnet2}.
The extractor was trained on VoxCeleb2~\cite{Chung2018} and used without benchmark-specific training.
The extractor maps each variable-duration speech segment to a single 192-dimensional embedding through its attentive statistics pooling head~\cite{Okabe2018}.
For each synthetic condition and contamination budget, we pooled metric counts over all development episodes, selected a single threshold, and then applied it unchanged to the evaluation split.
All cross-condition comparisons are \emph{matched-budget}: each condition uses a threshold selected on the development split at the same budget.

\begin{figure}[t]
\centering
\begin{subfigure}{0.5\linewidth}
\centering
\includegraphics[width=0.82\linewidth]{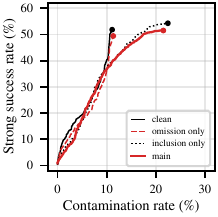}
\caption{Synthetic evaluation set}
\label{fig:curves-syn}
\end{subfigure}\hfill
\begin{subfigure}{0.5\linewidth}
\centering
\includegraphics[width=0.82\linewidth]{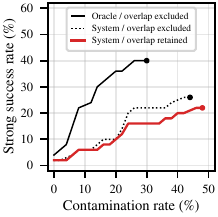}
\caption{Real evaluation set}
\label{fig:curves-real}
\end{subfigure}
\caption{Evaluation-set contamination--success curves.
Each curve is traced by sweeping the decision threshold; lowering it enrolls more names and generally raises both strong success and contamination.
Dots mark the no-threshold limit (\emph{saturation}).
Tables~\ref{tab:synth} and~\ref{tab:real} report fixed operating points along these curves.
(a)~Synthetic attendance-list conditions.
(b)~Real front-end diagnostic.}
\label{fig:curves}
\end{figure}

\begin{table}[t]
\centering
\caption{Synthetic evaluation-set results: each row is a development-split contamination budget used to select that condition's threshold.
Values are strong success\,/\,contamination (\%).
Budget 0 selects a threshold with zero observed contamination on the development split, and \emph{saturation} is the operating point obtained by removing the association-score threshold while keeping all other fixed abstention and eligibility rules.}
\label{tab:synth}
\footnotesize
\setlength{\tabcolsep}{3.5pt}
\begin{tabular}{lcccc}
\toprule
Contam.\ budget & Main & Clean & Incl.\ only & Omis.\ only \\
\midrule
0 & 1.00 / 0.00 & 2.13 / 0.14 & 0.76 / 0.05 & 3.62 / 0.34 \\
1\% & 1.95 / 0.24 & 7.04 / 0.71 & 3.83 / 0.76 & 9.03 / 1.74 \\
2\% & 6.53 / 1.86 & 9.13 / 1.09 & 7.52 / 1.75 & 10.81 / 2.36 \\
5\% & 15.63 / 4.53 & 19.15 / 4.87 & 16.64 / 4.87 & 17.95 / 5.21 \\
10\% & 35.16 / 9.34 & 31.87 / 8.65 & 38.35 / 10.45 & 39.62 / 10.38 \\
20\% & 50.83 / 18.82 & 51.82 / 11.11 & 53.95 / 21.47 & 49.42 / 11.34 \\
saturation & 51.50 / 21.53 & 51.82 / 11.11 & 54.28 / 22.46 & 49.42 / 11.34 \\
\bottomrule
\end{tabular}

\vspace{15pt}
\caption{Synthetic-to-real threshold-transfer diagnostic (strong success\,/\,contamination, \%).
Synthetic-evaluation results are repeated from Table~\ref{tab:synth}; real refit ($^{*}$) selects the threshold on the real evaluation labels and is not held out.
$^{\dagger}$~One-sided 95\% upper confidence bound for zero observed contamination under target independence: 5.82\%.}
\label{tab:real}
\footnotesize
\setlength{\tabcolsep}{4pt}
\begin{tabular}{lccc}
\toprule
Contam.\ budget & Synth.\ eval. & Real transfer & Real refit$^{*}$ \\
\midrule
0 & 1.00 / 0.00 & 0.00 / 0.00$^{\dagger}$ & 4.00 / 0.00 \\
10\% & 35.16 / 9.34 & 8.00 / 4.00 & 22.00 / 8.00 \\
20\% & 50.83 / 18.82 & 26.00 / 14.00 & 36.00 / 20.00 \\
saturation & 51.50 / 21.53 & 40.00 / 30.00 & 40.00 / 30.00 \\
\bottomrule
\end{tabular}
\vspace{5pt}
\end{table}

\textbf{Benchmark results.}
Fig.~\ref{fig:curves-syn} and Table~\ref{tab:synth} summarize performance on the synthetic evaluation set.
In the main condition at the 10\% budget, the method reaches $35.16\%$/$9.34\%$ strong success/contamination.
Relative to the clean condition, the main condition yields lower strong success at conservative operating points and higher contamination at the 20\% budget.
Clean lists remove attendance noise but not the cross-session association problem, and the $m{=}3$ gate leaves $37.1\%$ of the clean population unreachable (targets listed in fewer than three sessions); the clean saturation performance reflects coverage limits and residual association errors, not benchmark noise.

\textbf{Effects of attendance noise.}
Inclusion noise adds unsupported name--session associations.
At the task level, these dilute the evidence linking a name to a recurring voice and raise the risk of contaminated enrollment.
Under the reference method, spurious listings also raise a name's observed list frequency, allowing infrequent names to pass the $m{=}3$ eligibility gate.
This reduces the method-induced unreachable rate from 37.1\% under clean lists to 23.2\%.
Inclusion-dominated lists therefore primarily increase contamination risk.

Omission noise, by contrast, removes genuine name--session associations.
At the task level, this weakens a speaker's cross-session speech-presence evidence.
Under the reference method, it can also drop speakers below the $m{=}3$ gate or leave them unreachable, giving a method-induced unreachable rate of 39.0\%.
Consequently, omission primarily lowers saturation strong success, while saturation contamination remains close to that under clean lists.
Inclusion raises the observed saturation strong success, whereas omission lowers it (Table~\ref{tab:synth}).
The main-over-clean difference at the 10\% budget reflects separately selected thresholds and slightly different populations, not improved identifiability.

\textbf{Synthetic-to-real threshold transfer.}
Table~\ref{tab:real} reports performance under three settings: synthetic evaluation, fixed-threshold transfer to the real set, and diagnostic refitting on the real labels.
Fixed transfer applies the synthetic-development thresholds without adjustment, whereas real-set refitting reselects the threshold on the real labels and is therefore not held out.
Refitting recovers part of the loss at conservative budgets, but the resulting performance remains below that on the synthetic evaluation set.
Transfer and refitting coincide at saturation because threshold recalibration cannot affect this operating point.
At saturation, the $26.00\%$ method-induced unreachable rate and residual association or representation errors on real speech further limit performance.
The synthetic-to-real gap may also reflect acoustic differences---a known source of embedding and calibration mismatch~\cite{silnova22_odyssey}---since the benchmark is calibrated to attendance structure rather than the real corpus's acoustics.

Finally, to gauge the cost of the oracle front-end assumption, we evaluate a diarization system based on a high-resolution embedding extractor (HEE)~\cite{Heo2023a}, which reduces saturation strong success by about one third relative to oracle clustering (Fig.~\ref{fig:curves-real}).
This loss bundles segment-boundary, under-clustering, and over-clustering errors, and grows further when overlapping speech is retained.

\section{CONCLUSION}
\label{sec:conclusion}

We formalized multi-participant automatic enrollment---extracting reliable named enrollment speech from anonymous speaker clusters by cross-referencing them with noisy expected attendance lists.
To support this, we introduced a contamination-sensitive evaluation protocol and a synthetic benchmark calibrated to an in-house corpus.
The cross-session reference method revealed distinct effects of the two noise types.
However, reliable enrollment remained difficult at low contamination, and transferring the synthetic-development operating points to the real corpus revealed a substantial synthetic-to-real gap that refitting closed only partially, not at saturation.
Real-world validation rests on a single, private in-house corpus; we will release the generator, manifests, configurations, and evaluation code to support further work on system-diarization robustness and downstream recognition.

\clearpage
\bibliographystyle{IEEEbib}
\bibliography{strings,aid_local,papers}

\end{document}